# Near 100% CO Selectivity in Nanoscaled Iron-Based Oxygen Carriers for Chemical Looping Methane Partial Oxidation

Yan LIU[a†], Lang QIN[a†], Zhuo CHENG[a†], Josh W. GOETZE[a], Fanhe KONG[a], Jonathan A. FAN[b], Liang-Shih FAN[a*]

[a] Department of Chemical and Biomolecular Engineering, the Ohio State University, 151 W Woodruff Ave, Columbus, OH 43210, USA.

[b] Department of Electrical Engineering, Ginzton Laboratory, Spilker Engineering and Applied Sciences, 348 Via Pueblo Mall, Stanford University, Stanford, CA 94305, USA.

[†] Co-first authors

[*] Corresponding Author, fan.1@osu.edu

## Abstract

Chemical looping methane partial oxidation provides an energy and cost effective route for methane utilization. However, there is considerable $CO_2$ co-production in state-of-the-art chemical looping systems, rendering a decreased productivity in value-added fuels or chemicals. In this work, we show that the co-production of $CO_2$ can be dramatically suppressed in methane partial oxidation reactions using iron oxide nanoparticles, with a size of 5±3 nm, as the oxygen carrier. To stabilize these nanoparticles at high temperatures, they are embedded in an ordered, gas-permeable mesoporous silica matrix. We experimentally obtained near 100% CO selectivity in a cyclic redox system at 750 °C to 935 °C, which is a significantly lower temperature range than in conventional oxygen carrier systems. Density functional theory calculations elucidate the origins for such selectivity and reveal that $CH_4$ adsorption energies decrease with increasing nanoparticle size. These calculations also show that low-coordinated lattice oxygen atoms on the surface of nanoparticles significantly promote Fe-O bond cleavage and CO formation. We envision that embedded nanostructured oxygen carriers have the potential to serve as a general materials platform for achieving 100% selectivity in redox reactions at high temperatures.

Syngas, i.e., CO and $H_2$, is an important intermediate for producing fuels and value-added chemicals from methane via Fischer-Tropsch or other synthesis techniques[1]. Syngas has been produced commercially by steam reforming, auto-thermal reforming, and partial oxidation of methane for many decades[2]. However, an improvement of its energy consumption, environmental impact, and associated production cost has always been of interest. This has prompted the investigation into alternative routes that can avoid the use of air separation units for producing purified oxygen and are more effective in $CO_2$ emission control. It is also of interest to reduce the operating temperature of these processes, which are generally endothermic and traditionally require temperatures of 900 ℃ or higher to attain high reactant conversion rates. The use of high temperatures is problematic because the thermodynamic driving force for carbon deposition, and thus materials obliteration, can be accelerated[3]. Current approaches to reducing reaction temperatures while avoiding side product formation require noble metals such as Pt, Pd or Au, which leads to dramatic increases in cost[4].

Chemical looping methane partial oxidation[5] (CLPO) is an emerging approach that overcomes the above-mentioned shortcomings for syngas production. A CLPO process involves redox reactions taking place in two interconnected reactors: a reducer (or fuel reactor) and an oxidizer (also referred to as air reactor), shown in Figure 1(a). In contrast to conventional fossil fuel gasification and reforming processes, CLPO eliminates the need for an air separation unit, water–gas shift reactor, and acid gas removal unit. It has the potential to directly produce high-quality syngas with desirable $H_2$:CO ratios. The core of CLPO using natural gas as the feedstock involves complex redox reactions in which methane molecules adsorb and dissociate on metal oxide oxygen carrier surfaces. It also involves internal lattice oxygen ion diffusion in which oxygen vacancy creation and annihilation occurs. These reactions can be engineered to withstand thousands of redox cycles[5].

The recent progress in chemical looping technology for partial oxidation has advanced to the stage where successful pilot operation has been demonstrated and proven to be highly efficient with a minimal energy penalty in the process applications[6]. In this technology, the highest CO selectivity that the CLPO can reach is thermodynamically limited to 90% with accompanied 10% $CO_2$ generation. It is recognized that the $CO_2$ reduction in low purity syngas is extremely challenging and energy consuming, as $CO_2$ is among the most chemically stable carbon-based molecules[7]. This 10% of $CO_2$ in syngas can significantly reduce the productivity of value-added fuels or chemicals generated. Breaking away from the 90% CO selectivity limit to achieve a higher selectivity in redox reactions requires a different consideration from the metal oxide materials design and synthesis perspective. A promising approach is the use of metal oxide nanoparticles since the functionality of metal oxide materials intimately correlates with their structure in terms of size. Alalwan et al. investigated a set of unsupported α-$Fe_2O_3$ nanoparticles to explore the influence of particle size on $CH_4$ chemical looping combustion, and found decreasing the particle size from 350 to 3 nm increased the duration over which $CH_4$ was completely converted to $CO_2$. However, the unsupported nanoparticles underwent agglomeration and structural changes during the redox cycles, and as a result, they exhibited a steep decline in performance due to the loss of nanostructure.[8]

In the present work, we report a new approach to metal oxide oxygen carrier engineering for CLPO by designing and synthesizing nanoscale iron oxide carriers embedded in mesoporous silica SBA-15 ($Fe_2O_3$@SBA-15). Mesoporous silica is an engineered nanomaterial that has high surface area, ordered pore structures and high tunability of morphology. Its unique properties have attracted broad attention in a number of applications such as environmental treatment, catalysis and biomedical engineering[9-17]. SBA-15 is a common mesoporous silica that has perpendicular nanochannels with a narrow pore size distribution which is suitable for nanoparticle separation and gas penetration. A schematic of our materials platform is outlined in Figure 1. We experimentally achieve a high CO selectivity >99%, which is by far the highest in CLPO systems. We also find that cyclic methane partial oxidation with nanoscale oxygen carrier materials can be performed with high selectivity at temperatures as low as 750 °C. These findings underscore the strong size-dependent effect of metal oxide oxygen carriers at the nanoscale on syngas selectivity and reactant conversion in redox processes. This work will have broader impacts not only on CLPO, but also on other chemical looping applications such as carbonaceous fuel conversion and utilization.

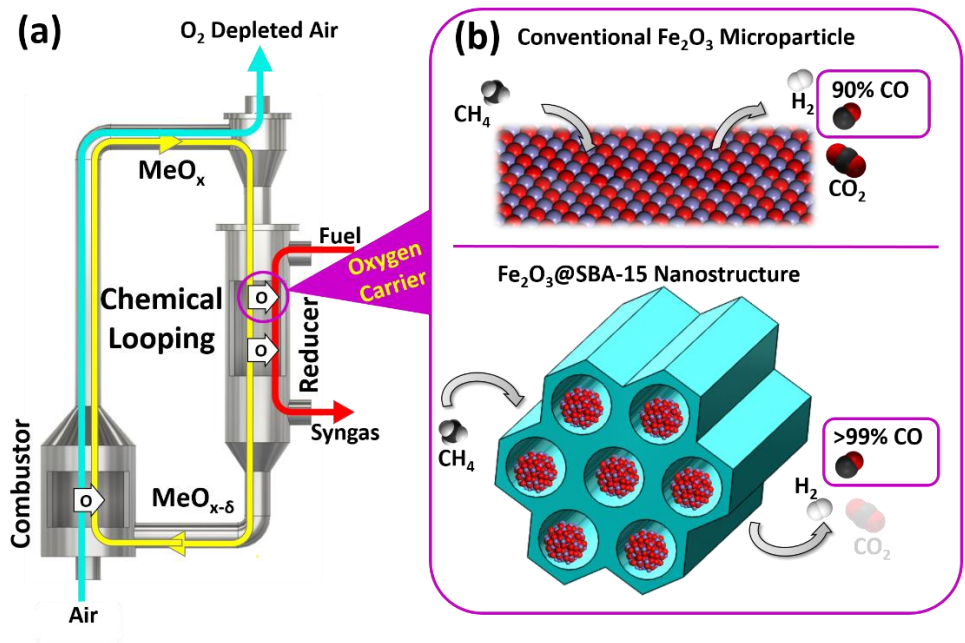

Figure 1 (a) Schematic of the chemical looping partial oxidation process, (b) structure and CO selectivity in conventional oxygen carrier vs $Fe_2O_3$@SBA-15 oxygen carrier.

**Results and Discussion**

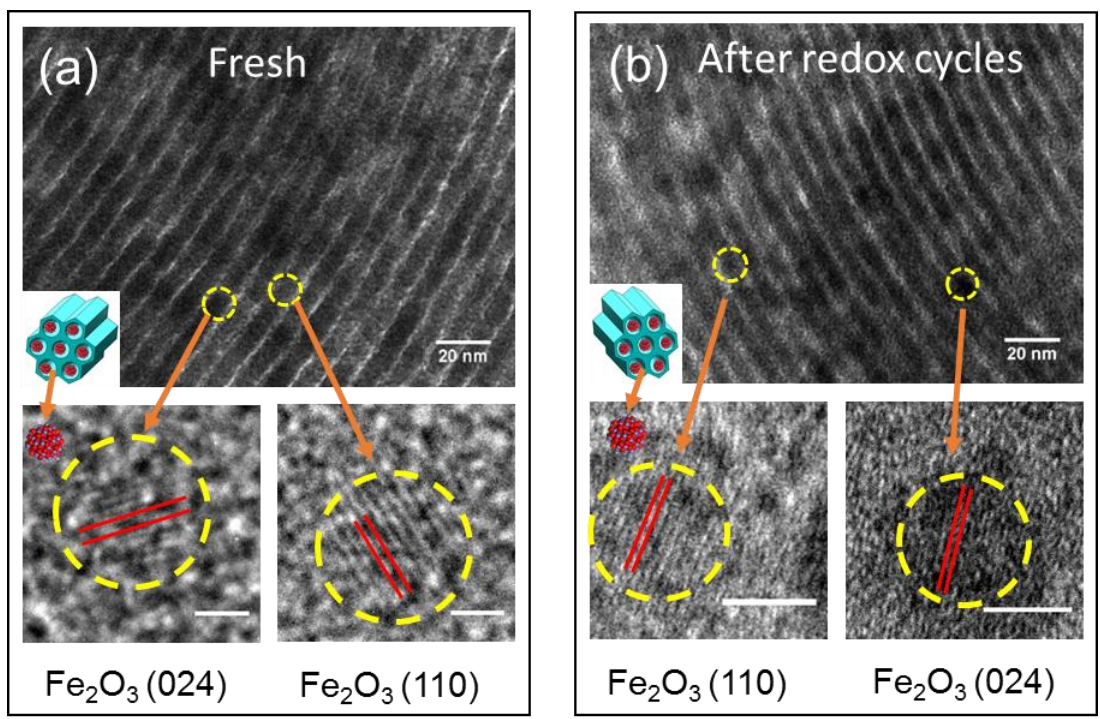

Figure 2. TEM images of (a) fresh $Fe_2O_3$@SBA-15 and HR-TEM images of two typical $Fe_2O_3$ nanoparticles (Scale bar represents 1nm); (b) $Fe_2O_3$@SBA-15 after multiple redox cycles and HR-TEM images of two typical $Fe_2O_3$ nanoparticles (Scale bar represents 5nm).

Figure 2 shows the TEM images of $Fe_2O_3$@SBA-15 before and after redox cycles. Nanoparticles with a size of 3-5 nm can be identified as $\alpha$-$Fe_2O_3$ based on lattice fringes measurement in which a = b = 5.038 Å and c = 13.772 Å. The images confirm that the nanoparticles remain embedded in SBA-15 nanochannels with identical morphology after multiple redox cycles. The particle size slightly increases to 5-8 nm after 75 redox cycles. This result indicates the high stability of $Fe_2O_3$@SBA-15 at high temperatures. The $Fe_2O_3$ particle size was also confirmed by $N_2$ physisorption and small angle X-ray diffraction (SAXRD), as shown in Supplementary Figure 1 and Supplementary Figure 2.

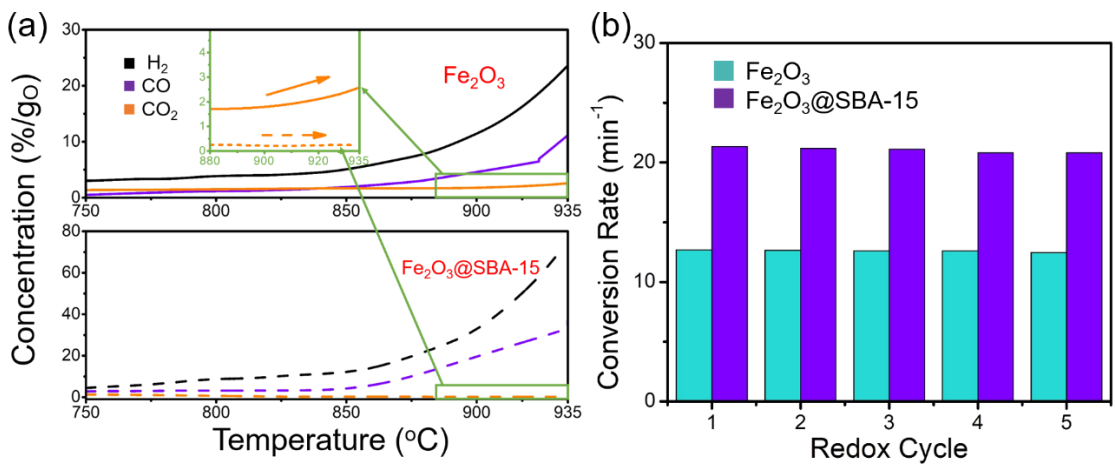

Figure 3. (a) Temperature ramped reduction results of $Fe_2O_3$@SBA-15 and $Fe_2O_3$; (b) Conversion rate during redox at 800°C

The reactivity and recyclability test results are shown in Figure 3. As shown in Figure 3a, a stable $CO_2$ concentration of <0.7%/$g_O$ is observed in $Fe_2O_3$@SBA-15 throughout methane partial oxidation, indicating a high syngas selectivity higher than 99.3%. In $Fe_2O_3$ without SBA-15 support, $CO_2$ formation is observed to increase with temperature, resulting in an average selectivity less than 87%. In the range of 750-935 °C, CO concentration from $Fe_2O_3$@SBA-15 is over 200% higher than $Fe_2O_3$ without SBA-15 support, suggesting a near 100% CO selectivity with significantly higher CO conversion rate.

Over 75 continuous redox cycles were carried out on both $Fe_2O_3$ samples with and without SBA-15 support. Five typical TGA redox cycles are shown in Figure 3b. The high recyclability is consistent with TEM observation. $Fe_2O_3$@SBA-15 not only has a high conversion rate, but is stable at high temperatures. This suggests that the separation of nanoparticles is essential in maintaining high CO selectivity, reactivity and recyclability. At 800°C, the conversion rate of $Fe_2O_3$@SBA-15 is 67% higher than $Fe_2O_3$ without SBA-15 support. The morphology of post redox nanoparticles can be found in Figure 2b, where $Fe_2O_3$@SBA-15

nanochannels remain almost identical to fresh samples. This demonstrates the high stability and anti-sintering effect of dispersed nanostructures at high temperatures.

To gain mechanistic insight into the role of the nanostructures in CO selectivity enhancement of $Fe_2O_3$@SBA-15, first-principles calculations were performed within the framework of density functional theory (DFT) using the Vienna Ab Initio Simulation Package (VASP)[18-20]. Figure 4 shows calculated energies of $CH_4$ adsorption on Fe atop site and O atop site of $(Fe_2O_3)_n$ nanoparticles as a function of n. The data of the previous computational study on $CH_4$ adsorption on $Fe_2O_3$ (001) surface are given by the filled circle. It can be seen that $CH_4$ adsorption energies dramatically decrease with increasing number, $n$ when the sizes of $Fe_2O_3$ nanoparticles are at a relatively small scale. However, they decrease slowly with increasing $n$ when the sizes are at relatively large scale. The strongest adsorption on $(Fe_2O_3)_4$ is $CH_4$ binding at the Fe atop site with an adsorption energy of 66.2 kJ/mol. The second stable configuration is $CH_4$ adsorption at the O atop site of $(Fe_2O_3)_4$ with an adsorption energy of 35.1 kJ/mol. When n increases from 4 to 60, the Fe atop adsorption becomes weaker with 43.9 kJ/mol lower adsorption energy. However, the adsorption at the Fe atop site and the O atop site of $(Fe_2O_3)_{60}$ nanoparticles is still stronger than adsorption on $Fe_2O_3$ (001) surface, as shown in Figure 5. This is because the average coordination number of surface Fe atoms in $(Fe_2O_3)_n$ nanoparticle is smaller than that on $Fe_2O_3$ (001) surface. The undercoordination results in an upward shift of the Fe $d$-band, yielding high binding energies.

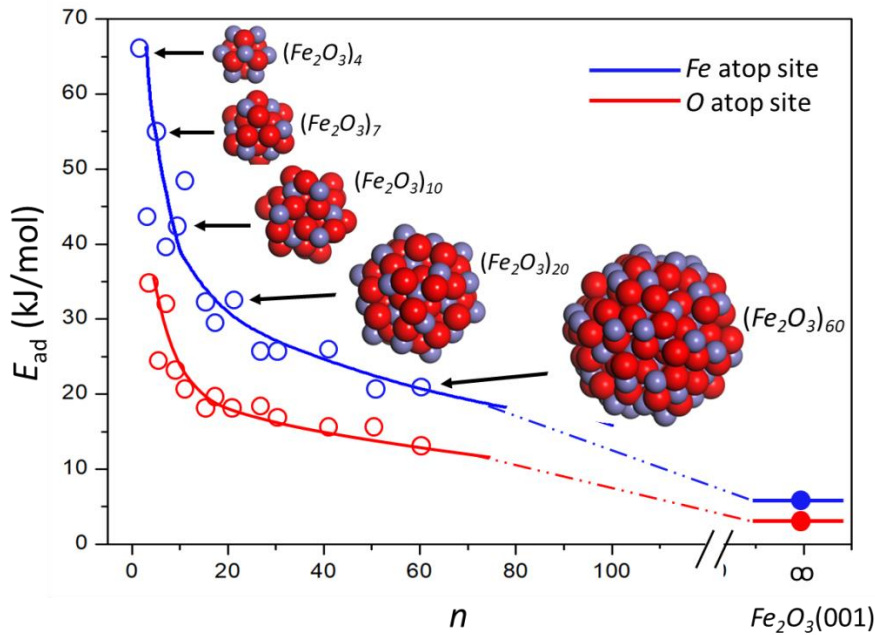

Figure 4. Calculated energies of $CH_4$ adsorption, $E_{ad}$ (open circles, kJ/mol), on Fe atop site and O atop site of $(Fe_2O_3)_n$ nanoparticles as a function of $n$. The adsorption trends are shown by the solid blue and red lines.

After methane activation, C-H bonds are cleaved to generate a carbon atom and four hydrogen atoms. To determine the dominant pathway for converting the carbon atom to CO on $(Fe_2O_3)_n$, a relatively small nanoparticle $Fe_{40}O_{60}$ ($n$=20) was chosen as the model to calculate the reaction barriers. $Fe_{40}O_{60}$ has three chemically distinguishable types of lattice oxygen atoms: 2-fold coordinated lattice oxygen $O_{2C}$, 3-fold coordinated lattice oxygen $O_{3C}$, and 4-fold coordinated lattice oxygen $O_{sub}$. As such, there are three reaction pathways for CO formation as shown in Figure 5(a). The calculated barriers for $CO_{2C}$, $CO_{3C}$ and $CO_{sub}$ formation are 37.2 kJ/mol, 69.6 kJ/mol and 58.5 kJ/mol, respectively. This result indicates C binding to $O_{2C}$ is the most favorable path, compared to C binding to $O_{3C}$ and $O_{sub}$ because Fe-O bonds of low-coordinated lattice oxygen atoms are easier to break than high-coordinated lattice oxygen atoms. In contrast to $Fe_{40}O_{60}$, all lattice oxygen atoms in the topmost atomic layer of the $Fe_2O_3$ (001) surface are three-coordinated atoms. Thus, the carbon atom on the $Fe_2O_3$ (001) surface converts to CO only via binding to $O_{3C}$, leading to a relatively high barrier of 61.2 kJ/mol as shown in Figure 5(b).

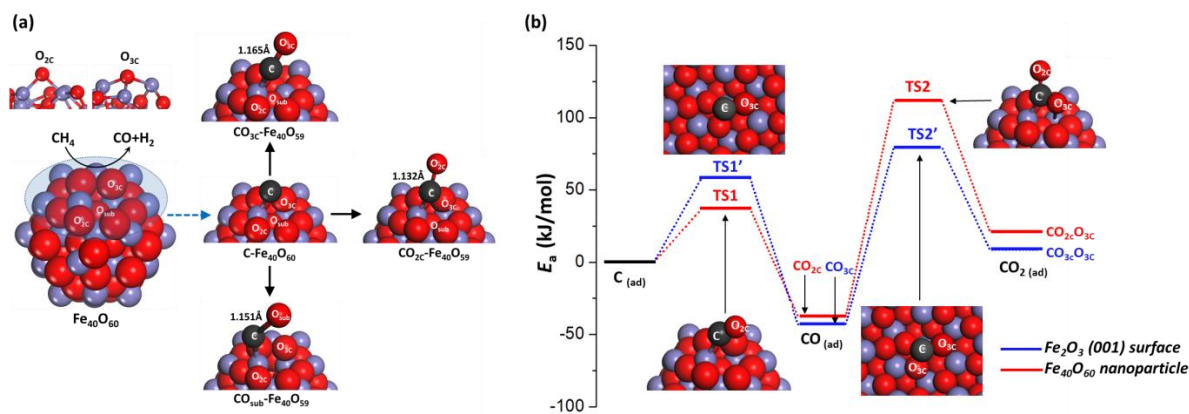

Figure 5. (a) Determining the pathway of CO formation on $Fe_{40}O_{60}$ nanoparticle. The C-O bond length and bond order are indicated. (b) Calculated barriers of CO and $CO_2$ formation on $Fe_{40}O_{60}$ (n=20) nanoparticles and $Fe_2O_3$ (001) surface.

The formed CO may further react with surface lattice O atoms to form $CO_2$.[21,22] For the $Fe_{40}O_{60}$ nanoparticle, the formation of $CO_2$ needs to overcome a barrier of 148.9 kJ/mol, which is 30.4 kJ/mol higher than that of $CO_2$ formation on $Fe_2O_3$ (001) surface. The high barrier with respect to $CO_2$ formation on $Fe_{40}O_{60}$ is attributed to the surface stress of nanoparticles, induced by surface atoms with unsaturated coordination. The surface stress leads to shorter and thus stronger Fe-$O_{3c}$ bonds compared to Fe-$O_{3c}$ bonds of the $Fe_2O_3$ (001) surface. The formation of $CO_2$ on $Fe_{40}O_{60}$ is endothermic, with the calculated reaction energy of 58.2 kJ/mol. These results indicate that the $CO_2$ formation on $Fe_{40}O_{60}$ is both kinetically and thermochemically unfavorable. Therefore, $Fe_{40}O_{60}$ nanoparticles significantly promote CO formation while inhibiting $CO_2$ production. Unfortunately, DFT-based calculations of large nanoparticles consisting of a few thousand atoms, required for confirming this conclusion, are intractable to compute even on the most powerful supercomputers. Nevertheless, the experimental evidence for the extraordinary CO selectivity of $Fe_2O_3$ nanoparticles with the size of 5±3 nm indicates that nanostructuring makes $Fe_2O_3$ a more active oxide for CO production compared to bulk $Fe_2O_3$.

In summary, we demonstrate that $Fe_2O_3$ nanoparticles embedded in SBA-15 enables near 100% CO selectivity in chemical looping methane partial oxidation, which is so far the highest value in product selectivity observed for chemical looping systems. Moreover, the effective temperature for syngas generation is lowered to 750-935 ℃, which is over 100 ℃ lower than current state-of-the-art processes. Nanostructured oxygen carriers are presented to exhibit little high-temperature reactivity property deterioration and adaptability to broader temperature operating windows for chemical looping operation conditions. These are important factors that can contribute to significant energy-saving reactor designs. The theoretical model and calculations reveal that the structure of the nanoparticles play a key role in CO selectivity enhancement of $Fe_2O_3$@SBA-15. The $CH_4$ adsorption energies and CO formation barriers depend not only on the nanoparticle size but also on the type of surface site exhibited by the nanoparticles. The small average coordination number of Fe atoms in the nanoparticle facilitates $CH_4$ adsorption and activation due to an upward shift of the Fe d-band, while the low-coordinated O atoms greatly promote Fe-O bond cleavage and CO formation, leading to a significant increase in CO selectivity. These findings contribute to a nanoscale understanding of the underlying metal oxide redox chemistry for chemical looping processes, and provide a systematic strategy toward the design of robust oxygen carrier nanoparticles with superior activity and selectivity.

**Author Contributes**

[†]Y.L., Q.L., and Z.C. contributed equally to this paper.

**Competing Interests**

The authors declare no competing interests.